\documentclass[12pt]{article}
\usepackage{graphicx}
\begin{document}

\title{The invisible hand and the rational agent are behind bubbles and crashes}
\author{Serge Galam\thanks{serge.galam@sciencespo.fr}, \\CEVIPOF, Centre de recherches politiques, \\Sciences Po and CNRS\\
 98, rue de l'Universit\'e, 75007 Paris, France}

\date{}
\maketitle

\begin{abstract}
The substantial turmoil created by both 2000 dot-com crash and 2008 subprime crisis has fueled the belief that the two classical paradigms of economics, which are the invisible hand and the rational agent, are not appropriate to describe market dynamics and should be abandoned at the benefit of alternative new theoretical concepts. At odd with such a view, using a simple model of choice dynamics from sociophysics, the invisible hand and the rational agent paradigms are given a new legitimacy. Indeed, it is sufficient to introduce the holding of a few intermediate mini market aggregations by agents sharing their own private information, to recenter the invisible hand and the rational agent at the heart of market self regulation including the making of bubbles and their subsequent crashes. In so doing, an elasticity is discovered in the market efficiency mechanism due to the existence of agents anticipation. This elasticity is found to create spontaneous bubbles, which are rationally founded, and at the same time, it provokes crashes when the limit of elasticity is reached. Although the findings disclose a path to put an end to the bubble-crash phenomena, it is argued to be rationality not feasible.
 \end{abstract}


\newpage

\section{The invisible hand and the rational agent}

In the last decades several crises in the financial world with subsequent heavy damages in the labor market and economy growth have put at stake the two classical paradigms of economics which are the invisible hand, the rational agent and their driving to associated equilibrium states \cite{as} . Indeed, the substantial turmoil created by the 2000 dot-com crash and 2008 subprime crisis have shaken quite many economists and financial analysts leading them to believe that those classical two paradigms should be abandoned at the benefit of alternative new theoretical concepts. At the heart of the questioning is the fundamental incapacity of neoclassical theory to embody the formation of bubbles and their following crashes, stating that equilibrium is always prevailing thanks to the existence of precisely 
the invisible hand and the rational agent. Before the recent 2000 and 2008 crashes a good deal of works had been performed to study bubbles formation and their bursts by physicists \cite{1, 2, 3, 4} as well by economists  \cite{5, 6, 7, 8, 9,10,11,12}.

At odd with such innovative prevailing views, we present a model inspired from sociophysics \cite{oliv, eco, review, santo, socio}, which produces bubbles as equilibrium states of a given market and crashes as the emergence of a new equilibrium. The model is rooted in neoclassical economy combined with the Galam model of opinion dynamics as the underline mechanisms leading to the step by step aggregation of individual choices towards the final collective equilibrium state. Accordingly, the invisible hand and the rational agent paradigms are shown to be indeed responsible in the making of market bubbles as well as in the associated crashes. Indeed the existence of an elasticity, together with a limit of the possible amplitude, in the market efficiency is uncovered driven by the making of local market efficiencies implemented by agent rationality.

To substantiate above surprising claim, it is enough to enlarge the perimeter of agent rationality to incorporate as self-interest the confrontation of oneÕs own actual decision to those of some other peer agents on the same market or asset restoring the concept of market self regulation. 

The main hypothesizes and the underlying model of individual choice aggregation are presented in next Section. The model is solved in third Section while the fourth Section introduces the existence of an elasticity in the efficiency of the market with respect reaching the fundamental values providing a theoretical basis to Keynes statement about the possible departure of stock prices from their fundamental values \cite{key1, key2}. The instrumental role of anticipation is emphasized in the process of bottom up aggregation of agent individual choices. The limit of elasticity is evaluated in Section fifth and last Section contains some concluding remarks.

\section{Model and main hypothesizes}

At the heart of market behavior in addition to the economical reality  stands human behavior and to study human behavior stands sociophysics. What is sociophysics ? It is the use of concepts and techniques from Statistical Physics to describe some social and political behaviors. It does not aim at an exact description  of the reality but at singling out some basic  mechanics which may be rather counter intuitive. Initiated more 33 years ago \cite{h1,h2}, it has started to become a main stream of research only in the last decade \cite{oliv, eco, review, santo, socio}.

Sociophysics deals with a rather large spectrum of problems including group decision making, coalition forming, terrorism, hierarchical voting, networks, linguistic, religion spreading, evolution, and finance should be included. One main focus is opinion dynamics, where models looks for generic mechanisms, which can be at work for a series of different public problems as Social, Political, Ecological, Societal, Economical, Behavioral, Innovation, Smoking, Rumors, Marketing, and Financial \cite{iori,y1,y2,j,walter}.

The background of the model is a bare frame of bimodal opinion dynamics in which agents are defined as rational. A rational agent has an opinion and advocates for it.  However, although a rational agent has a well grounded opinion, it is aware that the information it has access to is limited and may be misleading. Accordingly, it is susceptible to shift to the other opinion, if given more arguments for it. As it wants to make the best choice to optimize its profit, a rational agent does confront its current choice to the choices of other agents chosen from its social network. We consider that given a group of agents checking their mutual choices, each one advocating for its own choice, they end up following the local majority of initial choices. Therefore a rational agent updates its opinion by following the majority opinion from a group of selected agents including its own opinion. The update process produces a local polarization. Since, we have no access to the details of each agent discussions, we assume the groups are formed randomly.

However, within above framework, given an even size group, considering one agent one vote, a local tie may occur, with as many arguments in favor of buying a given asses as in favor of selling it. At this stage we make the additional hypothesis that the group of agents at a tie decides to lift the associated doubt aligning along the leading anticipation trend among them. If two agents are selling and two are buying, within a shared positive anticipation, the two sellers shift to two buyers and vice-versa in case of a shared negative anticipation. The introduction of possible tie, which in turn creates a local collective doubt, which is eventually turned to a common choice along the leading shared anticipation, makes the model counter-intuitive and non-trivial. Putting the process of local updates in equation leads to a threshold dynamics with a tipping point, which may be located anywhere between zero and one, depending on the group size distribution and the average market anticipation. Above the tipping point the update process increases the value of the relevant quantity and below the tipping point, the quantity is decreased.
Our two main hypothesizes can be formulated as follows:
\begin{description}
\item[(i)] 
The fundamental value of an asset or a stock is not accessible at once directly. All the information required to access to it is scattered into pieces of information among all the agents, which thus have all individual incomplete data. Accordingly, the total aggregation of all those pieces of information, which is revealed indirectly in the market value at the opening, contains the position of the current price fixing with respect to the fundamental value. Based on their respective private information some agents reach the right choice of selling or buying while others reach the wrong decision of buying or selling. 

If the proportion of initial buyers $b_0$ (with a proportion $(1-b_0)$ of initial sellers) is larger than fifty percents, the current price is underpriced. As a net result of the excess of buyers over the sellers the price should go up validating the market efficiency. In contrast, if the proportion of initial buyers $b_0$ (with a proportion $(1-b_0)$ of initial sellers) is smaller than fifty percents, the current price is overpriced. As a net result of the deficit of buyers over the sellers the price should go down validating  again the market efficiency. 

\item[(ii)]
Once every agent came out with its initial choice, to buy or to sell, it wants to get a kind of extra-check by creating a mini market aggregating a few other agent choices. In case of a local majority in favor of buying or selling, the agent adopt the majority choice. However in case of a local tie with the same number of buyers and sellers, the agent adopt the choice  in adequacy with the current leading anticipation of the market about the given stock or asset. Every agent repeats this local market updating some number of times before the market closure. 

Common beliefs towards anticipation vary depending on the stock and the general perception about the near future of the market as a whole. For balanced anticipations, in case of a doubt, agents keep on their respective opinions. For positive anticipation, in case of a doubt, agents choose to buy. For negative anticipation, in case of a doubt, agents choose to sell.

\end{description}

\section{The model}
By definition a rational agent reaches the right decision, to buy or sell a given asset,upon the analysis of its individual private information combined with the common information available to everyone. However, its private information is piecemeal since the complete information which determines the actual corresponding fundamental value is not accessible at once to one person, being fragmented into many pieces of different contents and meanings. Accordingly some agents end up making the  wrong true decision with respect to the asset while indeed doing the right decision according to their incomplete information. The market thus happens as the mean to aggregate all agent individual choices to access the true choice.  Aggregating all buyers and sellers yields a net pressure on the price at time $t=0$ given by,
\begin{equation}
p_0\equiv b_0-(1-b_0)=2b_0-1,
\label{p0}
\end{equation}
which is a real number satisfying $-1\leq p_0\leq 1$ with  $b_0$ being the proportion of buyers at time $t=0$ ($0\leq b_0\leq 1$ ) and $(1-b_0)$ the proportion of sellers. A positive pressure $p_0\geq0$ means the current price is undervalued with respect to the fundamental value while a negative value $p_0\leq0$ means the current price is overvalued with respect to the fundamental value. 

Nevertheless, the signal given by the pressure $p_0$ is fuzzy since when $0<p_0<1$ the market message about the current price is only probabilistic. For instance a value $p_0=0.70$ indicates that there is a chance of $70\%$ that the price is undervalued and $30\%$ that it is overvalued. At this point the market efficiency comes into play using agents trading to transform the probabilistic mere aggregation into a clear cut signal which, at time $T$ is either buy $(p_T=1)$ or sell $(p_T=0)$ at $100\%$ certainty. In between $p_0$ and $p_T$ the market is volatile with agents buying and selling before stabilization is reached with a limit up or down, i.e., $(p_T=1)$ or $(p_T=0)$.

To monitor the transformation of the probabilistic signal $p_0$ into a deterministic signal $(p_T=1)$ or $(p_T=0)$, agents create a series of mini markets to confront their respective choices. To implement those mini-markets, agents gather into small groups to confront their respective choices, thus aggregating their private information to reach a new common rational choice. 

To visualize the process, without loosing in robustness and generality, let us assume for illustration and simplicity of the calculations, that agents gather by groups of 4 respectively to gather their private information to update their individual choices.  Given an initial proportion of buyers $b_0$ at time $t=0$, one round of local updates leads o a new proportion of buyers $b_1$ at time $t=1$,
\begin{equation}
b_1 = b_0^4+4 b_0^3 (1-b_0)+6 (1-k) b_0^2 (1-b_0)^2,
\label{b1}
\end{equation}
where local groups are formed randomly. Last term includes the tie case (2 buyers - 2 sellers). In case of a tie, all 4 agents are doubting collectively. At that moment  doubting agents shift the doubt aligning along the choice in tune with the leading collective anticipation. A generalized optimistic anticipation is associated to $k=0$, i.e., in case of a local doubt, agents are sharing a common confidence about the future market performance of the asset, and thus decide to buy instead of selling. On the opposite side, a full generalized pessimistic anticipation about the current asset or market leads the agents to choose the selling position in case of a local doubt. Heterogeneity among agent anticipations is accounted by having $0\leq k \leq 1$ as the average market anticipation between the two extreme cases of euphoria ($k=1$) and depression ($k=1$). The net pressure on the price shifts from $p_0=2b_0-1$ to $p_1=2b_1-1$. Rewriting Eq.(\ref{b1}) as $p_1$ as a function of $p_0$ yields,
\begin{equation}
p_1 = \frac{-4p_0^3+12p_0+3(1-2k)(1-p_0^2)^2}{8}.
\label{p1}
\end{equation}

\subsection{The fully optimistic anticipation}

\begin{figure}[h]
\begin{center}
\includegraphics*[width=10cm,height=8cm]{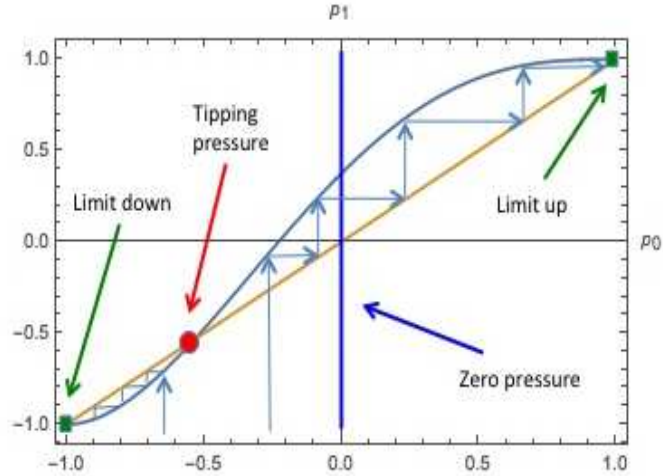}
\caption{The variation of $p_1$ as a function of $p_0$ from Eq. (\ref{p1}) within a fully optimistic anticipation ($k=0$). The market is deficient is the region $p_{c,0}<p_0<0$ and efficient for $p_0>0$ and $p_0<p_{c,0}$ with $p_{c,0}=\frac{2-\sqrt{13}}{3}\approx -0.535$.}
\label{fp1-0}
\end{center}
\end{figure}

Figure (\ref{fp1-0}) shows Eq. (\ref{p1}) for a full optimistic anticipation at $k=0$. It exhibits two attractors at $p_S=-1$ and $p_B=1$ separated by a threshold, a tipping point located at $p_{c,0}=\frac{2-\sqrt{13}}{3}\approx -0.535$. For $p_0<p_{c,0}$ the update yields $p_1<p_{c,0}$ while $p_1>p_{c,0}$ gives $p_1>p_0$. Since $p_{c,0}<0$ the update has reduced the signal fuzziness in the first case but increases it in the second case if $p_0<0$. Accordingly agents keep on updating to get a cleat cut signal with a deterministic choice shared by all agents within a limit choice, either up or down as seen from Figure (\ref{fp1-0}). At this stage, it is worth to notice from the Figure that between $p_{c,0}$ and the point at zero pressure, the pressure is negative and yet drives up the buyers proportion where it should naturally drives it down. In the area $p_{c,0}<p_0<0$ the market trend is deficient but as soon as the zero pressure point is passed by, the market recovers its efficiency with increasing positive pressures. On the other extreme, when $p_0<p_c$, the market is efficient at once, reducing till reaching the attractor $-1$.

\subsection{The fully pessimistic anticipation}

\begin{figure}[h]
\begin{center}
\includegraphics*[width=10cm,height=8cm]{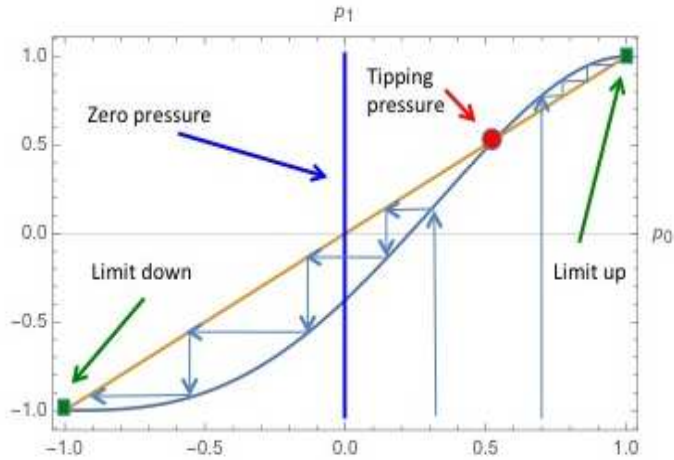}
\caption{The variation of $p_1$ as a function of $p_0$ from Eq. (\ref{p1}) within a fully optimistic anticipation ($k=1$). The market is deficient is the region $p_{c,1}<p_0<0$ and efficient for $p_0>0$ and $p_0<p_{c,1}$ with $p_{c,1}=\frac{-2+\sqrt{13}}{3}\approx 0.535$.}
\label{fp1-1}
\end{center}
\end{figure}

Figure (\ref{fp1-1}) shows Eq. (\ref{p1}) for a full pessimistic anticipation at $k=1$. It exhibits the same two attractors at $p_S=-1$ and $p_B=1$ as for the fully optimistic case but now the tipping point is located at a positive value $p_{c,1}=\frac{-2+\sqrt{13}}{3}\approx 0.535$. For $p_0<p_{c,1}$ the update yields $p_1<p_{c,1}$ while $p_1>p_{c,1}$ gives $p_1>p_0$. At contrast with the fully optimistic case, when $0<p_0<p_{c,1}$ the update increased the fuzziness by reducing a positive pressure till it cancels and only the once it is turns negative, does the market becomes efficient. For  $p_0>p_{c,1}$ the market is efficient at once as seen from Figure (\ref{fp1-1}). 

\subsection{The mixed anticipation}

Above two cases of a fully optimistic and pessimistic anticipations assumed all agents are sharing precisely the same anticipation at the same time. Combining those two extremes cases yields a mixed population with optimistic and pessimistic agents with an average anticipation $0\leq k\leq 1$ of the population, which accounts for a shaded anticipation as expressed in Eq. (\ref{p1})  \cite{hetero, deffu}. To determine the efficient/deficient regimes of the corresponding market, the fixed point equation $p_1=p_0$ is solved, yielding the same attractors $p_S=-1$ and $p_B=1$ separated by a threshold, a tipping point now located at,
\begin{equation}
p_{c,k}=\frac{-2+\sqrt{13-36k+36k^2}}{3(2k-1)} \ ,
\label{pck}
\end{equation}
with $p_{c,0}=\frac{2-\sqrt{13}}{3}\approx -0.535$ and  $p_{c,1}=\frac{-2+\sqrt{13}}{3}\approx 0.535$.

\begin{figure}[h]
\begin{center}
\includegraphics*[width=9cm,height=6cm]{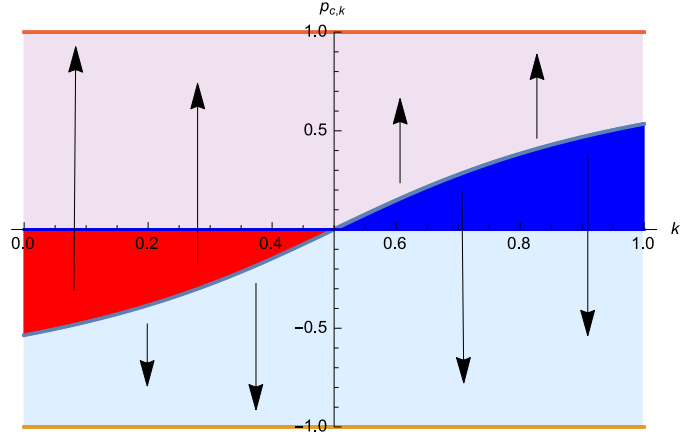}
\caption{The variation of the tipping point value  $p_{c,k}$ from Eq. (\ref{pck}) as a function of the average anticipation $k$. Dark areas around zero pressure, respectively negative and positive correspond to the elastic part of the market efficiency., i.e., where the market is deficient.}
\label{fpck}
\end{center}
\end{figure}

As seen from Figure (\ref{fpck}) the market is deficient in two symmetrical areas. First one occurs when $0\leq k <\frac{1}{2}$ for $p_{c,k}<p<0$ (red area), where the holding of the mini-markets drives a negative pressure toward zero value instead of heading to $p_S=-1$ before pushing it to positive pressures in the direction of a clear signal to buy at $p_B=1$. The reverse occurs when $\frac{1}{2}<k\leq 1$ for $0<p<p_{c,k}$ (dark blue area). Instead of getting an increase of the positive pressure, it is pushed down to zero before turning negative to eventually reach a clear signal to sell at $p_S=-1$.

\subsection{The balanced anticipation}

\begin{figure}[h]
\begin{center}
\includegraphics*[width=10cm,height=8cm]{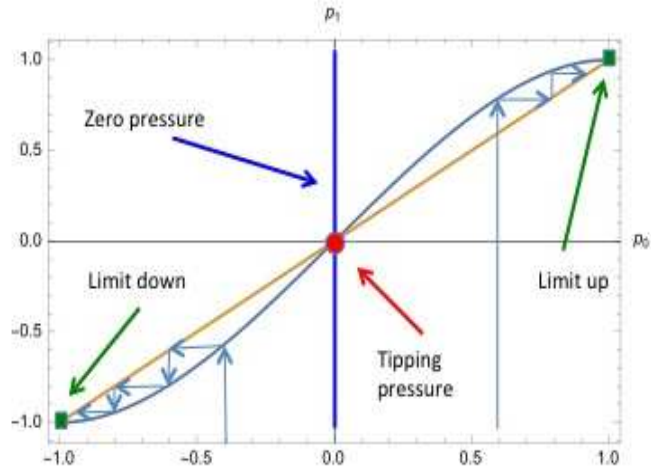}
\caption{The variation of $p_1$ as a function of $p_0$ from Eq. (\ref{p1}) within a balanced anticipation ($k=\frac{1}{2}$). The market is always efficient with no elasticity with $p_0<\frac{1}{2}\rightarrow p_S=-1$ and $p_0<\frac{1}{2}\rightarrow p_B=1$.}
\label{fp1-0.5}
\end{center}
\end{figure}

When anticipation is perfectly distributed among optimistic and pessimistic agents, the weighted value of the tie shift reaches $k=\frac{1}{2}$, which in turn makes the value of the tipping point undetermined with $p_{c,k}=\frac{0}{0}$ from Eq. (\ref{pck}). Going back to Eq. (\ref{p1}) yields for $k=\frac{1}{2}$
\begin{equation}
p_1 = \frac{-p_0^3+3p_0}{2},
\label{p1-1/2}
\end{equation}
which gives still the same attractors $p_S=-1$ and $p_B=1$ with a tipping point at $p_{c,\frac{1}{2}}=0$. Accordingly, the market is always efficient with no zones of deficiency as seen in  Figure (\ref{fp1-0.5}).

\section{How it works}

From last Section, the rationality of the agents combined with the market invisible hand results into the appearance of two regimes of market deficiency with the existence of an elasticity in the effect of market efficiency. While the aggregated private informations leads a buying or a selling trend, the current agents anticipation thwarts the efficiency to overcome the real signal reversing the trend direction. A bubble is formed. However this thwarting can be implemented only within a fixed range of the initial trend. The extend of elasticity is a function of the strength of the collectively shared anticipation. At some point the elasticity breaks down and the true trend is restored at once and brutally. It is a crash.

To be more explicit, consider an asset at time $t=0$ with a distribution of partial private informations yielding a pressure $p_0^1>0$ within a collective anticipation weighted by a value $k<\frac{1}{2}$ to shift a local mini-market tie into a buy choice, which yields a tipping point $p_{c,k}<0$. All external and internal conditions are assumed to be stable, i.e., nothing happens which could affect the current fundamental value.

Given those conditions, agents hold a series of $n$ successive mini-markets to transform $p_0^1$ to $p_1^1< p_2^1<p_3^1< ...<p_n^1=p_B=1$ making the initial fuzzy signal into a clear cut signal to buy. The market is efficient, thanks to the invisible hand and the agents rationality. The value of $n$ depends on both $p_0^1$ and $k$.

Once the signal is deterministic, corresponding prices go up. At that moment agents re-actualize their positions starting a second cycle of updates. Since the prices went up, the discrepancy from the fundamental value has been reduced and thus the new distribution of random private informations yields $0<p_0^2<p_0^1$. Anticipation is unchanged. 

Again, the invisible hand and the agents rationality combine to make the market efficient with the successive pressures $p_1^2<p_2^2<p_3^2< ...<p_m^2=p_B=1$ with a number $m$ larger than $n$ since the distance $p_0^2-p_B$ is larger than $p_0^1-p_B$.

The same process is iterated since at some number of cycle $l$ the initial fuzzy pressure $p_0^l$ gets negative with $p_0^l<0$. However, although the initial fuzzy pressure is negative and the invisible hand should drive it towards $p_S=-1$, since $p_0^l>p_{c,k}$, the efficiency elasticity produces a first series of negative pressures followed by positive pressures with $p_1^l<p_2^l< p_3^l<...<0<...<p_n^l=p_B=1$. A bubble formation is then initiated.

A new cycle follows with a negative pressure $p_0^{l+1}<p_0^l<0$ but still with $p_0^{l+1}>p_{c,k}$ within the elastic range of the market efficiency. Therefore, the pressure is pushed up towards a limit up trend. The bubble is formed. Additional cycles increases the number of updates to reverse the trend before reaching $p_B=0$ reinforcing the bubble.

However, without any notice, at some moment a new cycle $c$ is initiated at some value satisfying $p_0^{r}<p_{c,k}<0$ while all previous initial values satisfied  $p_0^{r}>p_{c,k}$. Having cross the elasticity limit, the market is efficient and the invisible hand pushes down the pressure till reaching the limit down clear signal at $p_S=-1$ in only a few updates since the distance $p_0^{r}-p_S$ is much smaller than $p_0^{r-1}-p_B$ of the precedent cycle. The bubble crashes.

\subsection{An illustration}

\begin{figure}[t]
\begin{center}
\includegraphics*[width=10cm,height=8cm]{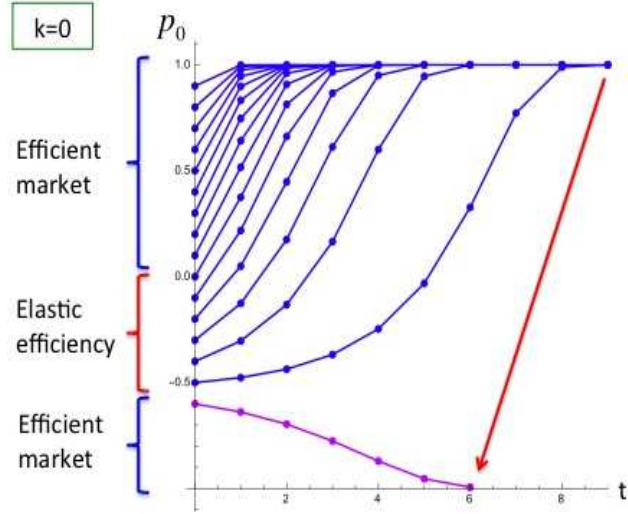}
\caption{Different cycles and associated steps to reach the clear signal in the case of a fully optimistic anticipation ($k=0$). Fifteen initial pressures from $p_0^1=0.90$ down to $p_0^{16}=-0.60$ by a decrement of $-0.10$ are shown. For the first 9 cycles ($p_0^1=0.90\rightarrow p_0^9=0.10$, the market is efficient. From  $p_0^{10}=0$ till $p_0^{15}=-0.50$ the market is deficient due to the optimistic anticipation. At $p_0^{16}=-0.60$, having cross the tipping point $p_{c,0}\approx -0.535$, the efficiency is restored brutally at once.}
\label{bubble1}
\end{center}
\end{figure}

Figure (\ref{bubble1}) shows above different cycles and steps in the case of a fully optimistic anticipation ($k=0$). The market starts at a higher pressure to buy at $p_0^1=0.90$, which means the current price is highly undervalued with $95\%$ of agents having the right private information. In one step, the signal to buy gets clear to everyone. Due to the buying, the price is corrected and at the next cycle, the pressure is at $p_0^2=0.80$ and 3 step are required to reach $p_B=1$. We artificially assume for the shake of illustration that every cycle produces a $-0.10$ correction to the initial pressure, i.e., $p_0^{j+1}=p_0^j-0.10$. The market has been fully efficient with a price reaching the fundamental value at $p_0^{10}=0$ after 9 successive cycles. At this stage, the market should have gotten stable with equal number of buying and selling agents.

However, due to the fully optimistic state of the market, the price keeps on changing and now reaches an overvalue leading to a next cycle with an initial negative value $p_0^{11}=-0.10$. But contrary to a normal local efficiency making, which should have lead to $p_S=-1$, the signal is thwarted to increase the pressure till reaching the wrong signal $p_B=1$. The only difference with the efficient earlier regime is the larger number of steps required to turn clear the signal. A bubble is born.

During the following cycles, a buy signal is again obtained increasing step by step the discrepancy between the fundamental value and the actual price, although the initial values $p_0$ get lower and lower reflecting those discrepancies.  However the anticipation thwarting effect cannot sustain a discrepancy with the fundamental value resulting into a negative pressure smaller than $p_{c,0}$. As soon as as $p_0<p_{c,0}\approx -0.535$, the efficiency is restored brutally at once provoking a simultaneous crash with an ending signal at $p_S=-1$. This breaking of elasticity occurs here at cycle 15 with $p_0^{16}=-0.60$. It is worth to stress that the number of steps is reduced from 9 for $p_0^{15}=-0.50$ to 6 for $p_0^{16}=-0.60$.

\begin{figure}[h]
\begin{center}
\includegraphics*[width=12cm,height=8cm]{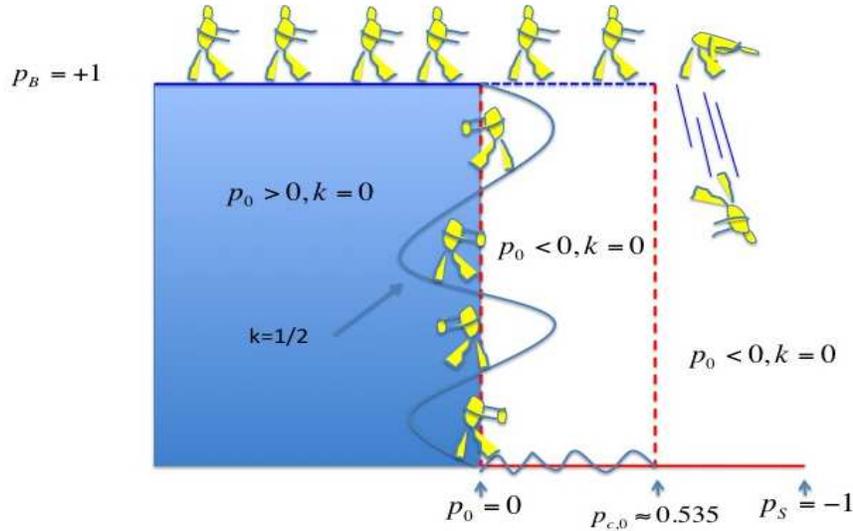}
\caption{The process of the anticipation thwarting of the efficiency mechanism in case of a positive anticipation ($k=0$). It is similar to cartoons (like Tom and Jerry) where the main character Tom the cat, runs after the mouse Jerry and at some point keeps on running in the void before realizing it is running in the void and then, at once falls down. At balanced anticipations ($k=\frac{1}{2}$), the net signal would move smoothly from buy ($p_B=+1$) to sell ($p_S=-1$). }
\label{bubble2}
\end{center}
\end{figure}

Figure (\ref{bubble2}) mimics the process of the anticipation thwarting of the efficiency mechanism in case of a positive anticipation similarly to cartoons (Tom and Jerry) where the main character Tom the cat, runs after the mouse Jerry and keeps running in the void before realizing it is running in the void and then, at once falls down.

\subsection{Anticipation collapse driven crashes}

Above illustration shows how the local market making, which implements the market efficiency does break momentally the efficiency within some range of private information, before recovering a full efficiency. The crash of a bubble is organically linked to its creation. All the process occurs given fixed external conditions. 

However, while in the elastic efficiency regime, the bubble can be crashed at once driven by a sudden change of the external conditions, provided those changes shift at once the anticipation from optimistic to pessimistic. Such a situation was implemented during the 2008 subprime mortgage crisis with the September 15 Lehman Brothers bankruptcy, which all of a sudden has project the collective anticipation into a fully pessimistic state. 

\begin{figure}[t]
\begin{center}
\includegraphics*[width=12cm,height=8cm]{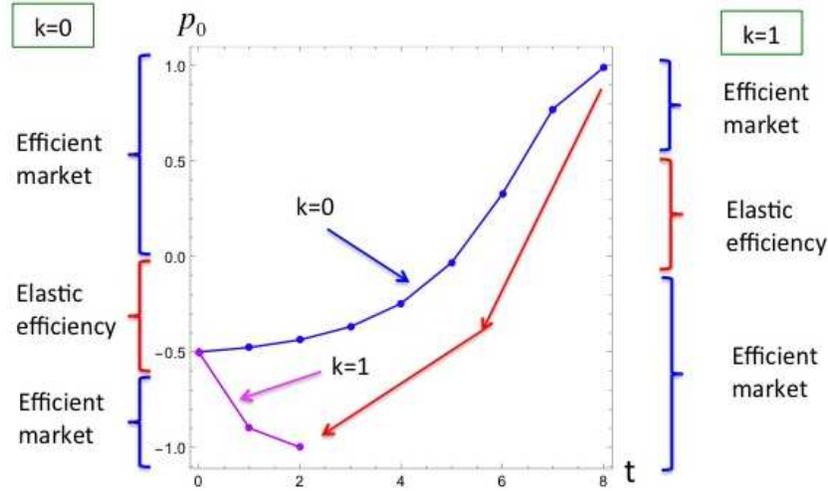}
\caption{An initial pressure $p_0=-0.50$ is shown with respectively an optimistic ($k=0$) and a pessimistic anticipation ($k=1$). In the first case, the local market making drives the signal towards $p_B=+1$ while in the second case the driving is towards $p_S=-1$. For $k=0$, being in the deficient zone, 8 steps are required to reach the thwarted clear signal but with $k=1$, being in an efficiency zone, the clear signal is reached within only 2 steps. }
\label{bubble3}
\end{center}
\end{figure}

An illustration is exhibited in Figure (\ref{bubble3}) with an initial pressure $p_0=-0.50$ with respectively an optimistic ($k=0$) and a pessimistic anticipation ($k=1$). In the first case, the local market making drives the signal towards $p_B=+1$ while in the second case the driving is towards $p_S=-1$. For $k=0$, being in the deficient zone, 8 steps are required to reach the thwarted clear signal but with $k=1$, being in an efficiency zone, the clear signal is reached within only 2 steps.
Therefore, jumping from an optimistic market into a pessimistic market, within an elastic efficiency zone, the signal is turned at once from $+1$ into $-1$ provoking a crash.

\section{Conclusion}

Applying the sociophysics  Galam model of opinion dynamics to marker dynamics we have been able to construct an efficient  market using an invisible hand to aggregate individuals private information to have the fundamental price reflected in the making of the current price. The process is built on through a series of sequential local market aggregations of private information to turn a fuzzy initial signal leading to a probabilistic distribution of individual choices, onto a deterministic clear signal to either buy or sell. It is worth to underline that the number of those micro markets steps is found to be a rather small number.

The rational agent hypothesis is shown to be instrumental in turning on the dynamics of market efficiency. However in so doing, due to the incompressible existence of agents anticipation, an elasticity has been disclosed in the efficiency dynamical process. It is this very elasticity, which on one hand produces spontaneous rational bubble, and on the other hand, crashes down those bubbles once the limit of elasticity is reached, breaking down the temporary deficiency. 

Sudden shifts of agent anticipations were also shown to create market crashes through sharp inversions of the efficiency elasticity range, which in turn produces sharp readjustments. 

Having identified the source of bubble-crash phenomena, we are in a position to determine a methodology to avert their occurrence. Indeed the efficiency elasticity should be suppressed to keep efficiency at a rigid status. Such a scheme requires to suppress the direction of anticipation about a given asset and or a market, which means to have $k=\frac{1}{2}$. Such a constraint can be implemented using two different paths. 

First path would demand to have on average an equal distribution of agents according to positive and negative anticipations. Such a constraint is hardly manageable in solid terms. 

Second path is easy to implement, at least in principle. It would require to have agents to toss a coin collectively at each case in which their local aggregated private information lead to a zero pressure choice, in order to make a choice to buy or to sell, and thus ignoring their own anticipation. 
From a rational stand, such a random practice would be be very hard to be followed psychologically since by nature people involved on a market are anticipating some outcome while making their respective choices. Moreover, if the practice is undertaken by part of the agents and not by the others, the elasticity will not be suppressed. 

It thus seem that bullish and perish dynamics are inherent to the existence of both a rational agent and an efficient market dynamics. Future work will include the study of having heterogeneity in the composition of agent behavior. In addition to rational agents, who make up their choices according to their private information rescaled by the local market aggregations, we will investigate the effect of stubborn and contrarian agents  \cite{contra, inflexe} on the elasticity range. 
Stubborn agents keep on their initial choice independently of outcomes of local market aggregations while contrarian agents decides against the local market outcomes.
 
\section*{Acknowledgment}

I would like to thank J. V. Andersen, M. Ausloos, S. Stefani and G. Rotundo for useful discussions about finance.
 


\newpage
\begin{thebibliography}{99}


\bibitem {as} A. Smith, The Theory of Moral Sentiments (1759, first edition), London: A Millar (1790)

\bibitem {1} T. Lux, Herd Behavior, Bubbles and Crashes, The Economic Journal 105 (1995) 881Ð896

\bibitem {2} N. Vandewalle and M. Ausloos, How the financial crash of October 1997 could have been predicted, The European Physical Journal 4 (1998) 139Ð141

\bibitem {3} R. N. Mantegna and H. E. Stanley, An Introduction to Econophysics: Correlations and Complexity in Finance, Cambridge University Press (1999)

\bibitem {4} D. Sornette, F. Deschatres, T. Gilbert and Y. Ageon, Endogeneous vs exogeneous shocks in complex systems: an empirical test using book sales ranking, Phys. Rev. Lett. 93 (2004) 228701 

\bibitem {5} O. Blanchard, Speculative Bubbles, Crashes and Rational Expectations, Economic Letters, 3 (1979) 387-389

\bibitem {6} C. Azariadis, Self-fulfilling prophecies, Journal of Economic Theory, vol. 25, (1981) 380-396
 
\bibitem {7}  R. Shiller, Stock Prices and Social Dynamics, Brookings Papers on Economic Activity, 2, (1984) 457-498 

\bibitem {8} B. Diba and H. Grossman, Explosive Rational Bubbles in Stock Price ?, American Economic Review, 78 (1988) 520-530

\bibitem {9} R. Shiller, Fashions, Fads, and Bubbles in Financial Markets, in  Coffee J., Ackerman S., Lowenstein L. (dir.), Knights, Raiders and Targets, Oxford University Press (1988) 56-68

\bibitem {10} A. Kirman, Epidemics of opinion and speculative bubbles in financial markets. in M. Taylor, Ed., Money and Financial Markets, Macmillan: London (1991)

\bibitem {11} A. Kirman, What or whom does the representative individual represent?, Journal of Economic Perspectives, 6 (1992) 117 

 \bibitem {12} A. Shleifer, Inefficient Markets : an Introduction to Behavorial Finance, Oxford University Press (2000)

\bibitem {oliv} De Oliveira S., De Oliveira P., Stauffer D. [1999], Non traditional Applications of computational statistical physics : Evolution, money, war and computers, Teubner, Stuttgart and Leipzig

\bibitem {eco} Econophysics and sociophysics: trends and perspectives, Bikas K. Chakrabarti, Anirban Chakraborti, Arnab Chatterjee (Eds.), Wiley-VCH (2006) 

\bibitem {review} S. Galam, Sociophysics: a review of Galam models, Int. J. Mod. Phys. C 19 (2008) 409440

\bibitem {santo} C. Castellano, S. Fortunato, V. Loreto, Statistical physics of social dynamics, Rev. Modern Phys. 81 (2009) 591Ð646

\bibitem {socio} S. Galam, Sociophysics, a physicistÕs modeling of psycho-political phenomena, Springer (2012)

\bibitem {key1} J. M. Keynes, A treatise on Probability, The Collected Writings of John Maynard Keynes, vol. VIII, London, Macmillan, 1973 (first print 1921)

\bibitem {key2} J. M. Keynes,The General Theory of Employment, Interest and Money, The Collected Writings of John Maynard Keynes, vol. VII, London, Macmillan, 1973 (first print 1936)

\bibitem {h1} S. Galam, ÒEntropie, dŽsordre et libertŽ individuelleÓ, Fundamenta Scientiae 3, 209-213 (1982)   

\bibitem {h2} S. Galam, Y. Gefen and Y. Shapir, ÒSociophysics: A mean behavior model for the process of strikeÓ, Journal of Mathematical Sociology 9, 1-13 (1982)

\bibitem {iori} G. Iori, A threshold model for stock return volatility and trading volume
International Journal of Theoretical and Applied Finance, Vol. 3, No. 3 (2000) 467-472

\bibitem {y1} Y. Biondi, and P.  Giannoccolo, Share Price Formation, Market Exuber- ance and Accounting Design, Banque de France Foundation Research Sem- inar, (2010¡ November 23

\bibitem {y2} Y.  Biondi, P. Giannoccolo and  S. Galam,  The formation of share market prices under heterogeneous beliefs and common knowledge, arXiv:1105.3228

\bibitem {j} J. Vitting Andersen, I. Vrontos, P. Dellaportas and S. Galam, Communication impacting financial markets, EPL, 108 (2014) 28007 (p1-p6)

\bibitem {walter} S. Galam, Valeur fondamentale et croyances collectives, in  Critique de la valeur fondamentale (French Edition) C. Walter and  E. Brian (Eds.), Chap. 5 (2008) 99-115

\bibitem{hetero} S. Galam, Heterogeneous beliefs, segregation, and extremism in the making of public opinions, Phys. Rev. E 71 (2005) 046123-1-5. 

\bibitem {deffu} G. Deffuant, D. Neau, F. Amblard, G. Weisbuch, Mixing beliefs among interacting agents, Adv. Complex Syst. 3 (2000) 87Ð98

\bibitem{contra} S. Galam, Contrarian deterministic effects on opinion dynamics, the hung elections scenario. Physica  A {\bf 333} (2004) 453

\bibitem {inflexe} S. Galam and F. Jacobs, The role of inflexible minorities in the breaking of democratic opinion dynamics, Physica A 381 (2007) 366376


\end{thebibliography}
\end{document}